\documentclass[english,reprint,amsmath,amssymb,aps]{revtex4-1}
\usepackage[latin9]{inputenc}
\usepackage{amsmath}
\usepackage{graphicx}
\usepackage{esint}

\makeatletter

\@ifundefined{textcolor}{}
{%
 \definecolor{BLACK}{gray}{0}
 \definecolor{WHITE}{gray}{1}
 \definecolor{RED}{rgb}{1,0,0}
 \definecolor{GREEN}{rgb}{0,1,0}
 \definecolor{BLUE}{rgb}{0,0,1}
 \definecolor{CYAN}{cmyk}{1,0,0,0}
 \definecolor{MAGENTA}{cmyk}{0,1,0,0}
 \definecolor{YELLOW}{cmyk}{0,0,1,0}
}

\usepackage{bm}
\usepackage[mathlines]{lineno}
\usepackage[english]{babel}
\usepackage[autostyle, english = american]{csquotes}
\MakeOuterQuote{"}

\makeatother

\usepackage{babel}
\begin{document}

\title{Heralded single phonon preparation, storage and readout in cavity optomechanics}

\author{Christophe Galland$^{1}$}

\email{chris.galland@epfl.ch}

\selectlanguage{english}%

\author{Nicolas Sangouard$^{2}$}

\author{Nicolas Piro$^{1}$}

\author{Nicolas Gisin$^{2}$}

\author{Tobias J. Kippenberg$^{1}$}

\affiliation{$^{1}$École Polytechnique Fédérale de Lausanne (EPFL), CH-1015 Lausanne,
Switzerland \\$^{2}$Group of Applied Physics, University of Geneva,
CH-1211 Genève 4, Switzerland}
\begin{abstract}
We show how to use the radiation pressure optomechanical coupling
between a mechanical oscillator and an optical cavity field to generate
in a heralded way a single quantum of mechanical motion (a Fock state).
Starting with the oscillator close to its ground state, a laser pumping
the upper motional sideband produces correlated photon-phonon pairs
via optomechanical parametric downconversion. Subsequent detection
of a single scattered Stokes photon projects the macroscopic oscillator
into a single-phonon Fock state. The non-classical nature of this
mechanical state can be demonstrated by applying a readout laser on
the lower sideband to map the phononic state to a photonic mode, and
performing an autocorrelation measurement. Our approach proves the
relevance of cavity optomechanics as an enabling quantum technology.
\end{abstract}
\maketitle

\paragraph*{Introduction}

\begin{figure}
\includegraphics{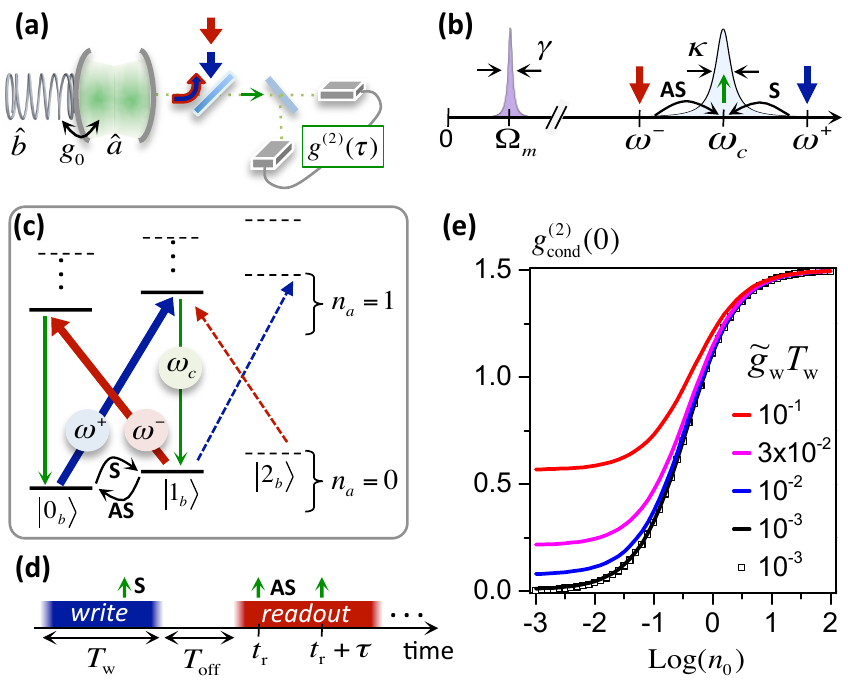} \protect\caption{\label{fig1}\textbf{(a)} Schematics of the optomechanical system
pumped at the upper and lower motional sidebands and of the correlation
measurement on the filtered cavity photons. \textbf{(b)} Representation
of the relevant mechanical and optical frequencies and linewidths.
\textbf{(c)} Resonant transitions during the \textit{write} and \textit{readout}
pulses. For a given mean photon number $n_{a}$ the number (Fock)
states of the mechanical oscillator form a harmonic ladder. Emission
at $\omega_{c}$ is enhanced by the cavity, which allows to address
selectively the Stokes, S (resp. anti-Stokes, AS) transitions between
phonon states when driving the upper (resp. lower) mechanical sideband.
\textbf{(d)} Pulse sequence (cooling step not shown). Detection of
a Stokes photon within the \textit{write} pulse duration $T_{\text{w}}$
is the heralding event. After a storage time $T_{\text{off}}$, coincidences
between anti-Stokes photons emitted at time $t_{\text{r}}$ and $t_{\text{r}}+\tau$
are measured. We define $t_{\text{r}}=0$ at the beginning of the
\textit{readout} pulse. \textbf{(e)} Conditional two-photon coincidence
as a function of the initial mechanical occupancy $\bar{n}_{0}$ (log
scale) for different values of the product $\widetilde{g}_{\text{w}}T_{\text{w}}$,
under the assumption of negligible mechanical damping ($T_{\text{w}}\ll(\gamma\bar{n}_{\text{th}})^{-1}$
and $T_{\text{off}}=0$). Open squares: only single-photon emission
events are post-selected,  eq.\eqref{g2cond}. Solid lines: including
the contribution from multiple photon emission. }
\end{figure}

Cavity optomechanical systems consist of a mechanical oscillator at
frequency $\Omega_{m}/2\pi$  coupled to an electromagnetic cavity
mode with resonant frequency $\omega_{c}/2\pi$ \cite{Aspelmeyer_2013}
(Fig.~\ref{fig1}a). The radiation pressure optomechanical coupling
can be used to either amplify \cite{Kippenberg_2008}  or cool \cite{Arcizet_2006,Schliesser_2006,Gigan_2006,WilsonRae_2007,Marquardt_2007,Schliesser_2009}
the mechanical degree of freedom. This has enabled the preparation
of mechanical oscillators in the quantum regime \cite{OConnell_2010,Teufel_2011,Chan_2011,Safavi_2012}
and the quantum coherent coupling between light and mechanical degrees
of freedom \cite{Verhagen_2012,Palomaki_2013}. Likewise, the optomechanical interaction allows for the
readout of mechanical motion with a readout imprecision below that at the standard
quantum limit \cite{Teufel_2009,Anetsberger_2010}.
In addition, optomechanically induced transparency \cite{Weis_2010}
can be utilized for slowing or advancing electromagnetic signals \cite{Safavi_2011,Zhou_2013}, for  coherent transfer between two optical wavelengths \cite{Hill_2012}, between the microwave and optical domains \cite{Bochmann_2013,Andrews_2013},
and for information storage and retrieval in long-lived oscillations
\cite{Palomaki_2013,Fiore_2011,McGee_2013}.

In the context of quantum information, \textit{continuous-variable}
schemes \cite{Schmidt_2012} such as optomechanical squeezing \cite{Vanner_2013,Szorkovszky_2013}
and entanglement \cite{Palomaki_2013b} in the quadrature operators
have been demonstrated in recent experiments. Yet there are many advantages
to using \textit{discrete} variables, for which heralded probabilistic
protocols can exhibit very high fidelity and loss-resilience \cite{Sangouard_2011}.
Moreover, on a fundamental level, studying quantized energy eigenstates
of macroscopic objects may allow new  tests of quantum mechanics \cite{Pepper_2012}
and of the nature of entanglement \cite{Pirandola_2006,Borkje_2011,Lee_2011}.
The first step toward this goal is to generate single-phonon Fock
states in long lived mechanical oscillators.

One possible route is to break the harmonicity of the system's eigenstates
by reaching the single-photon strong coupling regime \cite{Rabl_2011,Nunnenkamp_2011,Rips_2012,Ludwig_2012,Liao_2013,Qiu_2013,Kronwald_2013b,Rips_2013,Xu_2013b,Xu_2013c},
or to use the nonlinearity resulting from coupling to two level systems
\cite{Stannigel_2012,Ramos_2013}. However, the former requires $g_{0}/(\kappa,\varOmega_{m})\gtrsim1$,
where $g_{0}$ is the single-photon optomechanical coupling rate (see
below) and $\kappa$ is the total cavity energy decay rate $-$ a
regime far from state of the art experiments where $g_{0}/\kappa\sim10^{-3}$
\cite{Verhagen_2012,Chan_2012}. If multiple optical modes are introduced
a non conventional photon blockade regime can be used to relax the
constraint on the coupling strength \cite{Xu_2013,Savona_2013} $-$
and has been recently considered for conditional preparation of non-classical
states \cite{Basiri_2012,Komar_2013}. Projective measurements have also been proposed by Vanner \textit{et al.} to realize phonon addition and subtraction operations for general quantum state orthogonalization \cite{Vanner_2013b}.

In this Letter, we present an approach based on single-photon detection
to generate a single-phonon Fock state in a heralded way and then
convert it into a single photon, in the experimentally relevant \textit{weak-coupling}
and \textit{resolved-sideband} \cite{WilsonRae_2007,Marquardt_2007}
regime of a \textit{single-mode} optomechanical system (Figs.~\ref{fig1}a-d).
Starting with the mechanical mode close to its ground state (mean
phonon number $\bar{n}_{0}\ll1$), a \textit{write} laser pulse, tuned
to the upper motional sideband of the optical cavity, is used to amplify
\cite{Kippenberg_2005} the mechanical motion and generate (with low
probability) a correlated photon-phonon pair via optomechanical parametric
downconversion. The scattered photon $-$ referred to as \textit{Stokes}
photon in the following $-$ is spectrally-filtered from the pump
and detected by a photon counting module, thereby projecting the mechanical
oscillator (from its weak coherent state) into a single-phonon Fock
state while heralding the success of the procedure \cite{Vanner_2013b}. To verify the
non-classical nature of the heralded mechanical state, the mechanical
excitation is coherently mapped onto the optical cavity field by applying
a \textit{readout} laser tuned to the lower mechanical sideband (corresponding
to resolved sideband cooling \cite{Schliesser_2009}), and the statistics
of these \textit{Anti-Stokes} photons is analyzed in an autocorrelation
($g^{(2)}$) measurement \cite{Kimble_1977,Grangier_1986}. In the
limit where the \textit{write} (amplifying) and \textit{readout} (cooling)
pulses are shorter than the mechanical decoherence time, and for a
small enough initial phonon occupancy ($\bar{n}_{0}\ll1$), the two-fold
coincidence probability vanishes ($g^{(2)}\rightarrow0$) (Fig.~\ref{fig1}d),
demonstrating the heralded creation of a single-phonon Fock state
and its successful upconversion into a single cavity photon.

\paragraph*{Principle.}

We consider the optical and mechanical modes (represented by bosonic
operators $\hat{a}$ and $\hat{b}$, respectively) of an optomechanical
cavity driven by a laser on the lower or upper mechanical sideband,
corresponding to the angular frequencies $\omega_{\pm}=\omega_{c}\pm\Omega_{m}$
(Fig.~\ref{fig1}b). The Hamiltonian is a sum of three terms $\hat{\mathcal{H}}=\hat{\mathcal{H}}_{0}+\hat{\mathcal{H}}_{OM}+\hat{\mathcal{H}}_{dr\pm}$
describing the uncoupled systems, $\hat{\mathcal{H}}_{0}=\hbar\omega_{c}\hat{a}^{\dag}\hat{a}+\hbar\Omega_{m}\hat{b}^{\dag}\hat{b}$;
the optomechanical interaction, $\hat{\mathcal{H}}_{OM}=-\hbar g_{0}\hat{a}^{\dag}\hat{a}(\hat{b}^{\dag}+\hat{b})$;
and the laser driving, $\hat{\mathcal{H}}_{dr\pm}=\hbar(s_{\pm}^{*}e^{i\omega_{\pm}t}\hat{a}+s_{\pm}e^{-i\omega_{\pm}t}\hat{a}^{\dag})$,
where $\vert s_{\pm}\vert=\sqrt{\kappa P_{\pm}/\hbar\omega_{\pm}}$
is the incoming photon flux for a laser power $P_{\pm}$. As detailed
in \footnote[1]{See Supplemental Material below}, after switching to the interaction picture
with respect to $\hat{\mathcal{H}}_{0}$ and taking the weak-coupling
($g_{0}\ll\kappa$) and resolved-sideband ($\kappa\ll\Omega_{m}$)
limits we obtain the linearized Langevin equations during the \textit{write} (amplification) pulse
\begin{subequations}
\begin{align}
\frac{d\hat{a}}{dt} & =\frac{i}{\hbar}[\hat{\mathcal{H}}_{BSB},\hat{a}]-\frac{\kappa}{2}\hat{a}+\sqrt{\kappa}\hat{a}_{\text{in}}\label{Lang_a1}\\
\frac{d\hat{b}}{dt} & =\frac{i}{\hbar}[\hat{\mathcal{H}}_{BSB},\hat{b}]-\frac{\gamma}{2}\hat{b}+\sqrt{\gamma}\hat{b}_{\text{in}}\label{Lang_b1}
\end{align}
\end{subequations} with $\gamma$  the energy decay rate
of the mechanical oscillator. $\hat{\mathcal{H}}_{BSB}=-\hbar g_{+}\hat{a}^{\dag}\hat{b}^{\dag}+h.c$ is a parametric gain interaction and leads to the generation
of photon-phonon pairs (Fig.~\ref{fig1}c). Here $g_{+}=g_{0}\sqrt{\bar{n}_{\text{w}}}$
is the effective optomechanical coupling rate enhanced by the \textit{intracavity}
photon number $\bar{n}_{\text{w}}=\vert\frac{-is_{+}}{-i\Omega_{m}+\kappa/2}\vert^{2}=\frac{\kappa P_{+}}{\hbar\omega_{c}(\Omega_{m}^{2}+\kappa^{2}/4)}$
at the laser frequency. For simplicity we consider the optical
cavity to be overcoupled, i.e. the total cavity decay rate is dominated
by the external in/out-coupling rate $\kappa_{\text{ext}}$, so that
$\kappa\simeq\kappa_{\text{ext}}$. The operator $\hat{a}_{\text{in}}$
represents the vacuum noise entering the optical cavity, and $\hat{b}_{\text{in}}$
is the thermal noise from a phonon bath at temperature $T_{\text{bath}}$
and mean occupancy $\bar{n}_{\text{th}}\approx\frac{k_{B}T_{\text{bath}}}{\hbar\Omega_{m}}$.
The oscillator initial thermal occupancy $\bar{n}_{0}$ can be significantly
smaller than $\bar{n}_{th}$ if the \textit{readout} laser is also
used for sideband cooling (see Sec. III in SM) \cite{Schliesser_2009,Chan_2011}.

In a first simplified treatment, we neglect the decay of the mechanical
oscillator, which is a valid approximation if the pulse sequence is
shorter than the thermal decoherence time $(\gamma\bar{n}_{\text{th}})^{-1}$.
Since in our scheme $g_{+}\ll\kappa$, we can adiabatically eliminate
the cavity mode in eqs.(\ref{Lang_a1},\ref{Lang_b1}) $\hat{a}_{\text{w}}(t)\simeq\frac{2}{\kappa}(ig_{+}\hat{b}_{\text{w}}^{\dag}+\sqrt{\kappa}\hat{a}_{\text{w,in}})$.
Using the input/output relations \cite{Gardiner_1985} $\hat{a}_{\text{w,out}}=-\hat{a}_{\text{w,in}}+\sqrt{\kappa}\hat{a}_{\text{w}}$
(the subscript $\text{\text{w}}$ refers to the operators during the
\textit{write} pulse) we obtain the coupled optomechanical equations\begin{subequations}
\begin{align}
\hat{a}_{\text{w,out}} & =\hat{a}_{\text{w,in}}+i\sqrt{2\tilde{g}_{\text{w}}}\hat{b}_{\text{w}}^{\dag}\label{Lang_wa1}\\
\frac{d\hat{b}_{\text{w}}}{dt} & =\tilde{g}_{\text{w}}\hat{b}_{\text{w}}+i\sqrt{2\tilde{g}_{\text{w}}}\hat{a}_{\text{w,in}}^{\dag}\label{Lang_wb1}
\end{align}
\end{subequations}
where $\tilde{g}_{\text{w}}\equiv\frac{2g_{+}^{2}}{\kappa}$.
Introducing the temporal modes \cite{Hofer_2011} for the cavity driven
by a \textit{write} pulse of duration $T_{\text{w}}$, $\hat{A}_{\text{w,in/out}}(T_{\text{w}})=(\frac{\pm2\tilde{g}_{\text{w}}}{1-e^{\mp2\tilde{g}_{\text{w}}T_{\text{w}}}})^{-\frac{1}{2}}\int_{0}^{T_{\text{w}}}e^{\mp\tilde{g}_{\text{w}}t}\hat{a}_{\text{in/out}}(t)dt$,
we can write the solutions of Eqs.(\ref{Lang_wa1},\ref{Lang_wb1})
as $U^{\dag}A_{\text{w,in}}U$ and $U^{\dag}b_{\text{w}}(0)U$ where
the propagator $U$ is given by \footnote[1]

\begin{multline}
U(T_{\text{w}})=e^{i\sqrt{1-e^{-2\tilde{g}_{\text{w}}T_{\text{w}}}}\hat{A}_{\text{w,in}}^{\dag}\hat{b}_{\text{w}}^{\dag}}\\
e^{\tilde{g}_{\text{w}}T_{\text{w}}(-1-\hat{A}_{\text{w,in}}^{\dag}\hat{A}_{\text{w,in}}-\hat{b}_{\text{w}}^{\dag}\hat{b}_{\text{w}})}e^{-i\sqrt{1-e^{-2\tilde{g}_{\text{w}}T_{\text{w}}}}\hat{A}_{\text{w,in}}\hat{b}_{\text{w}}}
\end{multline}
For an oscillator initially in a thermal state characterized by the
density matrix $\rho_{b}(0)=(1-p)\sum_{n\geq0}p^{n}\vert n\rangle\langle n\vert$
with $p\equiv\frac{\bar{n}_{0}}{1+\bar{n}_{0}}$ the state of the
\emph{optomechanical} system at the end of the write pulse is $\rho_{A,b}(T_{\text{w}})=U(T_{\text{w}})(\vert0_{A}\rangle\langle0_{A}\vert\otimes\rho_{b}(0))U^{\dagger}(T_{\text{w}})$.
The conditional \emph{mechanical} state upon detection of a single
photon in mode $\hat{A}_{\text{w,out}}$ is obtained by applying the
projection operator $\vert1_{A}\rangle\langle1_{A}\vert\otimes\mathbb{I}_{b}$,
tracing out the optical mode and normalizing, 
\begin{align}
\rho_{b}^{\text{cond}}(T_{\text{w}}) & =\frac{tr_{A}\:\{(\vert1_{A}\rangle\langle1_{A}\vert\otimes\mathbb{I}_{b})\rho_{A,b}(T_{\text{w}})\}}{tr_{A,b}\:\{(\vert1_{A}\rangle\langle1_{A}\vert\otimes\mathbb{I}_{b})\rho_{A,b}(T_{\text{w}})\}}\nonumber \\
 & =(1-\overline{p})^{2}\sum_{n\geq0}\overline{p}^{n}(n+1)\vert n+1\rangle\langle n+1\vert\label{eq:rhocond}
\end{align}
where $\overline{p}=pe^{-\tilde{g}_{\text{w}}T_{\text{w}}}$. For
a small gain parameter ($\tilde{g}_{\text{w}}T_{\text{w}}\ll1$),
which is essential to maximize the probability of successful single-phonon
heralding (see \footnote[1])), and a resonator initially in its ground state
($p\ll1$), the dominant term is the single-phonon Fock state $\vert1_{b}\rangle$.

In the \textit{readout} step, driving the lower sideband at $\omega_{-}$
leads to the beam-splitter interaction $\hat{\mathcal{H}}_{RSB}=-\hbar g_{-}\hat{a}^{\dag}\hat{b}+h.c$ (with $g_{-}=g_{0}\sqrt{\bar{n}_{\text{r}}}$ and $\bar{n}_{\text{r}}$ the intracavity photon number at the red sideband) replacing $\hat{\mathcal{H}}_{BSB}$ in Eqs.(\ref{Lang_a1},\ref{Lang_b1}),
which coherently swaps the optical and mechanical states (Fig.~\ref{fig1}c).
The phonon statistics can thus be mapped onto the anti-Stokes photons
and subsequently be measured with a Hanbury-Brown Twiss setup (Fig.~\ref{fig1}a)
\cite{Kimble_1977}. Following similar steps as above, we compute
the zero-delay second-order autocorrelation of the anti-Stokes photons
during the \textit{readout} pulse, $g_{\text{cond}}^{(2)}(0)=\frac{\langle\hat{A}_{\text{r,out}}^{\dag2}\hat{A}_{\text{r,out}}^{2}\rangle_{c}}{\langle\hat{A}_{\text{r,out}}^{\dag}\hat{A}_{\text{r,out}}\rangle_{c}^{2}}$,
where the expectation value is taken on the post-selected mechanical
state, eq.\eqref{eq:rhocond}. We find 
\begin{equation}
g_{\text{cond}}^{(2)}(0)=\frac{2\overline{p}(2+\overline{p})}{(1+\overline{p})^{2}}\approx4\bar{n}_{0}\label{g2cond}
\end{equation}
where the last approximation is valid in the limit $\bar{n}_{0}\ll1$
and $\tilde{g}_{\text{w}}T_{\text{w}}\ll1$. This result shows that
the two-fold coincidence probability vanishes linearly with $\bar{n}_{0}$
and proves the non-classical nature of the heralded phonon state.
In Fig.~\ref{fig1}e we plot eq.\eqref{g2cond} along with the results
obtained when multiple photon emission is taken into account (see
\footnote[1])) for different values of the gain parameter $\tilde{g}_{\text{w}}T_{\text{w}}$.
We note that for sufficient \textit{readout} laser power the internal
phonon-to-photon conversion efficiency, approximated by $1-e^{-2\tilde{g}_{\text{r}}T_{\textrm{r}}}$
in the limit $p\ll1$ ($\tilde{g}_{\text{r}},T_{\textrm{r}}$ are
given explicitly in \footnote[1])), can be close to 1.

Let us briefly recall the conditions for observing strong antibunching:
(i) \textit{Weak-coupling} and \textit{resolved-sideband} regime:
$g_{0}\ll\kappa\ll\Omega_{m}$; (ii) Negligible mechanical decoherence:
$T_{\text{w}}+T_{\text{off}}\ll(\gamma\bar{n}_{\text{th}})^{-1}$;
and (iii) High initial occupancy of the ground-state: $\bar{n}_{0}\ll1$.
Because the pulse duration is bounded from below by $T_{\text{w}}>1/\kappa$
(the spectral width of the pulse should be narrower than the cavity),
we can recast (ii) onto the condition: $\gamma\bar{n}_{\text{th}}\ll\kappa$.
Noting that for a given bath temperature $\bar{n}_{\text{th}}\propto1/\Omega_{m}$,
this shows that the oscillator should have both a large $Q$ and a
large frequency $\Omega_{m}$.

\begin{figure}
\includegraphics{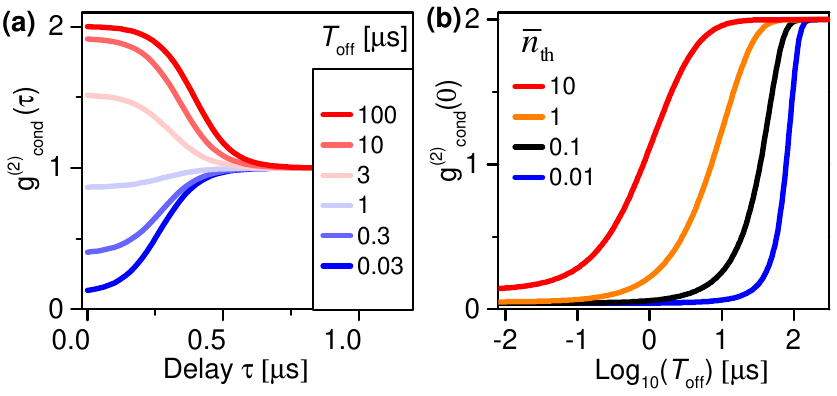} \protect\caption{\label{fig2}\textbf{(a)} Conditional second-order correlation function
$g_{\text{cond}}^{(2)}(\tau\vert t_{\text{r}})$ at fixed $t_{\text{r}}=1$~ns
for increasing waiting time $T_{\text{off}}$ between the \textit{write}
and \textit{readout} pulses. The bath temperature is set to 1.6~K
($\bar{n}_{\text{th}}=6.4$). The transition from antibunching to
bunching is a signature of the relaxation from a single-phonon Fock
state to a thermal state. \textbf{(b)} Same-time correlation $g_{\text{cond}}^{(2)}(0)$
(i.e. two-photon emission probability) as a function of $T_{\text{off}}$
for decreasing phonon bath thermal occupancy $n_{\text{th}}$, yielding
coherence times up to $(\gamma\bar{n}_{\text{th}})^{-1}\sim100$~$\mu$s. }
\end{figure}

\paragraph*{Experimental Feasibility.}

Many optomechanical systems have already been demonstrated that satisfy
(i) and for which condition (ii) would be trivially achieved owing
to the typically long mechanical decay time \cite{Goryachev_2012,Sun_2012,Chakram_2013}
, but condition (iii) is challenging to meet in these systems. Here
we consider a photonic crystal nanobeam resonator \cite{Kuramochi_2010,Chan_2011,Chan_2012},
for which the very high frequency of the confined phonon mode ($\Omega_{m}/2\pi\simeq5$
GHz) is beneficial. For a given bath temperature, fewer quanta are
thermally excited, while a large $\Omega_{m}$ also  facilitates spectral
filtering of the (anti-)Stokes photons from the pump laser beam (e.g.
with high-Finesse Fabry-Perot filters). Moreover, the structures reported
in \cite{Chan_2012} exhibit large optomechanical coupling rate $g_{0}/2\pi\simeq1$MHz
and their optical linewidth $\kappa/2\pi<1$GHz place them in the
resolved-sideband regime. Finally, coherence times of $\mathcal{O}(10-100\cdot10^{-6}\text{\text{{s}}})$
are within reach at 4~K and below \cite{Sun_2013} as mechanical
energy decay rates of $\gamma/2\pi\simeq7.5$~kHz have been measured
at 10~K \cite{Chan_2012}.

Using the  parameters reported in \cite{Chan_2012} and a realistic
bath temperature $T_{\text{bath}}\approx1.6$~K (corresponding to
He-4 buffer gas cooling \cite{Riviere_2013}), an initial occupancy
of $\bar{n}_{0}\approx0.01$ can be achieved by 100~ns of sideband
cooling with a peak intracavity photon number  $\bar{n}_{\text{r}}=10^{3}$
 corresponding to 150~$\mu$W of peak external laser power (see \footnote[1]),
Sec. III). The cooling laser is switched off during the \textit{write/store} sequence. Including mechanical dissipation, we integrate eqs.(\ref{Lang_a1},\ref{Lang_b1})
and compute $g_{\text{cond}}^{(2)}(\tau\vert t_{\text{r}})$, the
probability for anti-Stokes photon emission at times $t_{\text{r}}$
and $t_{\text{r}}+\tau$ during the \textit{readout} pulse, conditioned
on the detection of a herald photon during the \textit{write} pulse
(see Fig.~\ref{fig1}d). In Fig.~\ref{fig2}a we plot $g_{\text{cond}}^{(2)}(\tau\vert t_{\text{r}}=1$~ns)
for fixed \textit{write} pulse parameters $T_{\text{w}}=50$~ns and
$\bar{n}_{\text{w}}=0.1$, corresponding to a probability of Stokes
emission $\sim2\tilde{g}_{\text{w}}T_{\text{w}}(1+\bar{n}_{0})\sim10^{-3}$/pulse.
For waiting times between the \emph{write} and  \emph{readout} pulses
shorter than the decoherence time of the mechanics, $T_{\text{off}}\ll(\gamma\bar{n}_{\text{th}})^{-1}\simeq20$~$\mu$s,
we observe clear antibunching, a signature of successful conversion
of the phonon Fock state into a single photon.

Beyond verifying the non-classical state of the macroscopic oscillator,
our results also suggest a new tool for the on-demand generation of
single photons \cite{Chou_2004,Chen_2006,Matsukevich_2006}. Within
a time-window $\sim(\gamma\bar{n}_{\text{th}})^{-1}$ the heralded
Fock state is stored in the mechanical oscillator and can be retrieved
on-demand by applying the \textit{readout} pulse. 

Some advantageous features of the optomechanical systems considered
here is that the single photons are emitted in a well-defined spatial
mode and may be coupled into a single-mode fiber with high efficiency
$>90\%$ \cite{Cohen_2013,Groblacher_2013b}. Operation over the entire
electromagnetic wavelength range and integration into large scale
photonic circuits \cite{Streshinsky_2013} are other appealing assets. By engineering a cavity supporting two optical modes both coupled to a same mechanical mode, one could generate the herald photon and release the readout photon at two arbitrary wavelengths.
Although the \textit{write} step is intrinsically probabilistic, it
is possible to achieve near-deterministic Fock state creation by employing
simple feedback techniques \cite{Felinto_2006,Chen_2006,Matsukevich_2006}. 

\begin{figure}
\includegraphics{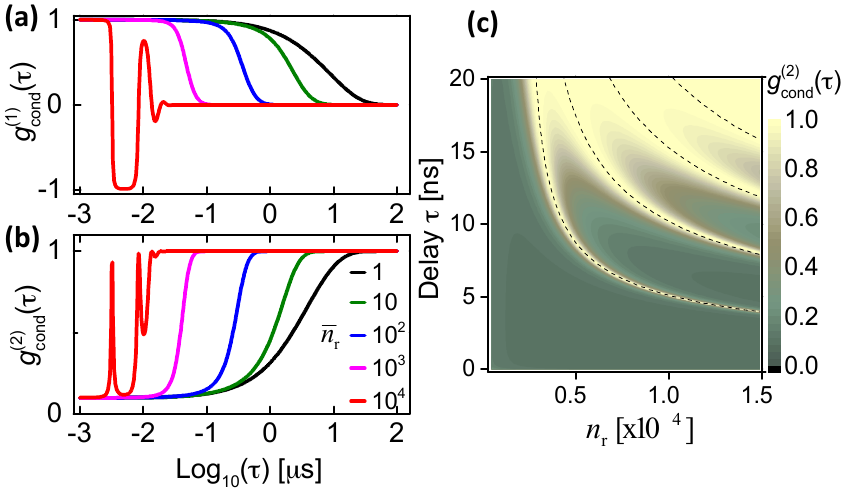} \protect\caption{\label{fig3}\textbf{Shaping single photons (a)} Field and \textbf{(b)}
intensity two-time correlation as a function of the time delay $\tau$
between photons, for increasing \textit{readout} pulse intracavity
photon number $\bar{n}_{\text{r}}$ (from black to red). Bath temperature
is 1.6~K and $T_{\text{off}}=5$~ns. The coherence time can be tuned
over three orders of magnitude from $10$~$\mu$s down to $10$~ns,
corresponding to linewidths from 100~kHz to 100~MHz. \textbf{(c)}
Color plot of $g_{\text{cond}}^{(2)}(\tau)$ versus $\bar{n}_{\text{r}}$
and $\tau$. Rabi oscillations set in when the laser-enhanced optomechanical
coupling becomes larger than the cavity decay rate: $g_{-}>\kappa$.
Dashed lines show the expected maxima for the Rabi period $\frac{1}{2}\sqrt{g_{0}^{2}\bar{n}_{\text{r}}-\frac{1}{16}\kappa^{2}}$.}
\end{figure}

Our scheme additionally enables precise control on the linewidth and
coherence properties of the on-demand single photons \cite{Almendros_2009}
by tuning the strength of the \textit{readout} pulse characterized
by the peak intracavity photon number $\bar{n}_{\text{r}}$ (at the
sideband $\omega_{-}$), as shown in Fig.~\ref{fig3}. In the limit
of weak \textit{readout} laser ($\bar{n}_{\text{r}}\lesssim1$) the
anti-Stokes photon coherence time is set by the thermal coherence
time of the oscillator $(\gamma\bar{n}_{\text{th}})^{-1}$. Increasing
$\bar{n}_{\text{r}}$ shortens the coherence time and eventually we
reach the (laser-enhanced) strong coupling regime $g_{-}\gtrsim\kappa$
and observe the onset of Rabi oscillations for $\bar{n}_{\text{r}}\gtrsim10^{4}$,
corresponding to multiple phonon-photon swapping cycles within the
optical cavity lifetime. This yields a remarkable range of achievable
coherence times, and therefore provides a way to generate on-demand
single photons with tunable linewidths from tens of kHz to hundreds
of MHz, an interesting feature for envisioned quantum networks, e.g.
to couple various physical realizations of nodes using photons as
carriers of quantum information.

\paragraph*{Entanglement and quantum repeaters.}

The potential applications of optomechanical systems become more evident
when noting the analogy with the scheme based on Raman transitions
in atomic ensembles first proposed by Duan \textit{et al.} \cite{Duan_2001}
to achieve scalable entanglement distribution between distant nodes
(DLCZ protocol). Specifically, consider two distant optomechanical
systems coherently excited by a weak laser beam, such that the probability
that both systems are simultaneously excited is negligible. The resulting
Stokes modes are interfered on a beamsplitter \cite{Borkje_2011}
and the detection of a single photon projects the distant mechanical
oscillators into an entangled state where they share a single delocalized
phonon. Successive entanglement swapping operations can then be used
to extend the entanglement over hundreds of kilometers \cite{Sangouard_2008}.

As a quantitative example, let us estimate the average time $T_{\text{ent}}$
required to establish entanglement between two optomechanical resonators
separated by 10~km of optical fiber using the DLCZ scheme. To first
order in the small parameter $\tilde{g}_{\text{w}}T_{\text{w}}$ we
have \cite{Sangouard_2011,Sekatski_2011}: $T_{\text{ent}}\simeq(2R_{r}\cdot2\tilde{g}_{\text{w}}T_{\text{w}}\cdot\eta)^{-1}$
where $R_{r}$ is the repetition rate of the experiment and $\eta$
the overall detection efficiency of the Stokes photons. For the particular
system considered here, realistic values are $R_{r}=10$~MHz and
$\eta=0.5\times0.6\times0.2=6\%$, where the three factors correspond,
in this order, to the collection in a single-mode fiber, the propagation
over 10~km of fibers, and the detection efficiency. Although $T_{\text{ent}}$
can be made shorter by increasing $\tilde{g}_{\text{w}}T_{\text{w}}$,
this also increases the probability for multiple pair excitation and
thereby decreases the fidelity expressed as $F\simeq(1-\bar{n}_{0})^{-1}(1-3\cdot2\tilde{g}_{\text{w}}T_{\text{w}}\cdot(1-\eta))$.
Assuming a target fidelity of $F=0.9$ \cite{Sangouard_2008} and
$1-\bar{n}_{0}\sim1$ we obtain $\tilde{g}_{\text{w}}T_{\text{w}}\simeq0.017$
and thus $T_{\text{ent}}\simeq23.5$~$\mu$s. Remarkably, this time
is slightly shorter than the light propagation time of $\sim50$~$\mu$s,
which would therefore set the lower bound on entanglement distribution
time.

In summary, we have shown how to generate a single-phonon Fock state
in an optomechanical resonator under the experimentally accessible
\textit{weak-coupling} and \textit{resolved-sideband} regimes. Starting
with the oscillator in its motional ground state, a \textit{write}
laser pulse tuned on the upper mechanical sideband creates correlated
phonon-photon pairs. The detection of the Stokes photon heralds the
successful preparation of a single-phonon Fock state in the mechanical
oscillator. Finally, the non-classical statistics of the phonon state
is mapped onto the optical field by a \textit{readout} pulse tuned
on the lower sideband, and conditional two-photon correlations reveal
antibunching. Our proposal opens promising perspectives for the use
of optomechanical systems as quantum memories and on-demand single-photon
sources for emerging applications in quantum information processing
and communication.
\begin{acknowledgments}
This work was financially supported by the EU project SIQS. The authors also ackowledge the Swiss National Science Foundation (SNSF)
for its support through the NCCR QSIT. C.G. is supported by an SNSF
\emph{Ambizione} Fellowship and N.P. acknowledges the support from
the FP7 Marie Curie Actions of the European Commission, via the IEF
fellowship QPOS (project ID 303029). C.G. and N.S. would like to thank
Vivishek Sudhir and Pavel Sekatski, respectively, for useful discussions.
\end{acknowledgments}

\bibliographystyle{apsrev4-1}
\bibliography{biblioQOM}

\pagebreak 

\newpage 

\section*{SUPPLEMENTAL MATERIAL}

\section{Calculations of correlation functions}
\label{Sec1}

\subsection{Linearized Langevin equations}

We consider an optomechanical system with a single relevant optical
mode (annihilation operator $\hat{a}$, frequency $\omega_{c}/2\pi$)
and a single relevant mechanical mode (annihilation operator $\hat{b}$,
frequency $\Omega_{m}/2\pi$). The cavity is driven by a laser tuned
either on the red or blue mechanical sideband, i.e. $\omega_{\pm}=\omega_{c}\pm\omega_{m}$.
The corresponding Hamiltonian is a sum of three terms 
\[
\hat{\mathcal{H}}=\hat{\mathcal{H}}_{0}+\hat{\mathcal{H}}_{OM}+\hat{\mathcal{H}}_{dr\pm}
\]
describing, respectively, the uncoupled systems, the optomechanical
interaction and the laser driving \begin{subequations}\label{Hamiltonians}
\begin{align}
\hat{\mathcal{H}}_{0} & =\hbar\omega_{c}\hat{a}^{\dag}\hat{a}+\hbar\Omega_{m}\hat{b}^{\dag}\hat{b}\label{H_0}\\
\hat{\mathcal{H}}_{OM} & =-\hbar g_{0}\hat{a}^{\dag}\hat{a}(\hat{b}^{\dag}+\hat{b})\label{H_om}\\
\hat{\mathcal{H}}_{dr\pm} & =\hbar(s_{\pm}^{*}e^{i\omega_{\pm}t}\hat{a}+s_{\pm}e^{-i\omega_{\pm}t}\hat{a}^{\dag})\label{H_dr}
\end{align}
\end{subequations}

The single-photon optomechanical coupling rate is $g_{0}/2\pi$ and
$\vert s_{\pm}\vert=\sqrt{\kappa P_{\pm}/\hbar\omega_{\pm}}$ is the
incoming photon flux for a laser power $P_{\pm}$ driving the cavity
at the higher/lower mechanical sideband. For simplicity we have considered
that the total cavity decay rate is dominated by the external coupling
rate $\kappa=\kappa_{\text{ext}}+\kappa_{\text{int}}\simeq\kappa_{\text{ext}}$.
We switch to the interaction picture with respect to $\hat{\mathcal{H}}_{0}$
by applying the unitary transformation $\hat{\mathcal{H}}\rightarrow e^{\frac{i}{\hbar}\hat{\mathcal{H}}_{0}t}\hat{\mathcal{H}}e^{-\frac{i}{\hbar}\hat{\mathcal{H}}_{0}t}$.
The optomechanical coupling and the driving terms are expressed in
this frame as

\begin{subequations}\label{Interaction} 
\begin{align}
\hat{\mathcal{H}}_{OM} & =-\hbar g_{0}\hat{a}^{\dag}\hat{a}(\hat{b}^{\dag}e^{i\Omega_{m}t}+\hat{b}e^{-i\Omega_{m}t})\label{H_om_int}\\
\hat{\mathcal{H}}_{dr\pm} & =\hbar(s_{\pm}^{*}e^{\pm i\Omega_{m}t}\hat{a}+s_{\pm}e^{\mp i\Omega_{m}t}\hat{a}^{\dag})\label{H_dr_int}
\end{align}
\end{subequations}

We write the Langevin equations (without the noise terms for now)
with the energy decay rates $\kappa/2\pi$ and $\gamma/2\pi$ for
optical and mechanical excitations, respectively, 
\begin{align*}
\frac{d\hat{a}}{dt} & =\frac{i}{\hbar}[\hat{\mathcal{H}},\hat{a}]-\frac{\kappa}{2}\hat{a}\\
 & =ig_{0}\hat{a}(\hat{b}^{\dag}e^{i\Omega_{m}t}+\hat{b}e^{-i\Omega_{m}t})-is_{\pm}e^{\mp i\Omega_{m}t}-\frac{\kappa}{2}\hat{a}\\
\frac{d\hat{b}}{dt} & =\frac{i}{\hbar}[\hat{\mathcal{H}},\hat{b}]-\frac{\gamma}{2}\hat{b}\\
 & =ig_{0}\hat{a}^{\dag}\hat{a}e^{i\Omega_{m}t}-\frac{\gamma}{2}\hat{b}
\end{align*}

Following Wooley and Clerk \cite{Wooley_2013} we make the Ansatz
$\hat{a}(t)=\hat{a}_{0}(t)+\hat{a}_{\pm}(t)e^{\mp i\Omega_{m}t}$
and $\hat{b}(t)=\hat{b}_{0}(t)$. Neglecting all terms rotating at
$\pm n\Omega_{m}$ for $n\geq2$, i.e. assuming the good cavity limit
$\kappa\ll\Omega_{m}$, we obtain a set of equations at the Fourier
frequencies $\omega=0,\pm\Omega_{m}$ 
\begin{subequations} 
\begin{align}
\frac{d\hat{a}_{0}}{dt} & =ig_{0}(\hat{a}_{-}\hat{b}_{0}+\hat{a}_{+}\hat{b}_{0}^{\dag})-\frac{\kappa}{2}\hat{a}_{0}\\
\frac{d\hat{b}_{0}}{dt} & =ig_{0}(\hat{a}_{0}^{\dag}\hat{a}_{+}+\hat{a}_{-}^{\dag}\hat{a}_{0})-\frac{\gamma}{2}\hat{b_{0}}\\
\frac{d\hat{a}_{+}}{dt} & =ig_{0}\hat{a}_{0}\hat{b}_{0}-is_{+}-(\frac{\kappa}{2}-i\Omega_{m})\hat{a}_{+}\label{a_plus}\\
\frac{d\hat{a}_{-}}{dt} & =ig_{0}\hat{a}_{0}\hat{b}_{0}^{\dag}-is_{-}-(\frac{\kappa}{2}+i\Omega_{m})\hat{a}_{-}\label{a_minus}
\end{align}
\end{subequations} 
We make the second approximation of weak single-photon
optomechanical coupling $g_{0}\ll\kappa,\Omega_{m}$ so that we can neglect the
nonlinear terms proportional to $g_{0}$ in eqs. (\ref{a_plus}-\ref{a_minus}).
Since we are interested in interaction times long compared to $\kappa^{-1}$
we can ignore the transient behaviors of the fields $\hat{a}_{\pm}$
and substitute their steady-state values

\[
\overline{a}_{+}\simeq\frac{-is_{+}}{-i\Omega_{m}+\kappa/2}\quad\text{ ; }\quad\overline{a}_{-}\simeq\frac{-is_{-}}{i\Omega_{m}+\kappa/2}
\]

We therefore arrive at the following linearized Langevin equations
(we drop the operator indices $0$ for simplicity), including the
input noise operators, \begin{subequations}\label{Langevin} 
\begin{align}
\frac{d\hat{a}}{dt} & =\frac{i}{\hbar}[\hat{\mathcal{H}}_{lin},\hat{a}]-\frac{\kappa}{2}\hat{a}+\sqrt{\kappa}\hat{a}_{\text{in}}\label{Lang_a}\\
\frac{d\hat{b}}{dt} & =\frac{i}{\hbar}[\hat{\mathcal{H}}_{lin},\hat{b}]-\frac{\gamma}{2}\hat{b}+\sqrt{\gamma}\hat{b}_{\text{in}}\label{Lang_b}
\end{align}
\end{subequations}

with the linearized Hamiltonian 
\begin{equation}
\hat{\mathcal{H}}_{lin}=-\hbar(g_{+}\hat{a}^{\dag}\hat{b}^{\dag}+g_{-}\hat{a}^{\dag}\hat{b})+h.c\label{LinearHamiltonianNoCR}
\end{equation}
where $g_{+}=g_{0}\overline{a}_{+}$ (resp. $g_{-}=g_{0}\overline{a}_{-}$)
is the effective optomechanical interaction rate enhanced by the intracavity
field. The intracavity photon number at the frequency of the \textit{write}
(resp. \textit{readout}) laser pulse is thus $\bar{n}_{\text{w}}=\vert\overline{a}_{+}\vert^{2}$
(resp. $\bar{n}_{r}=\vert\overline{a}_{-}\vert^{2}$). Without loss
of generality we can also take $g_{\pm}$ real since we are not interested
in interference effects that could arise were the two lasers simultaneously
driving the cavity.

The thermal (Markovian) noise entering the optical and mechanical
cavity modes is characterized by the operators $\hat{a}_{\text{in}}$
and $\hat{b}_{\text{in}}$, respectively. The non-zero second-order
moments of the noise operators are \begin{subequations}\label{NoiseG1}
\begin{align}
\langle\hat{a}_{\text{in}}(t)\hat{a}_{\text{in}}^{\dag}(t')\rangle & =\delta(t-t')\label{G1a_in}\\
\langle\hat{b}_{\text{in}}(t)\hat{b}_{\text{in}}^{\dag}(t')\rangle & =(\bar{n}_{\text{th}}+1)\delta(t-t')\label{G1b_in}\\
\langle\hat{b}_{\text{in}}^{\dag}(t)\hat{b}_{\text{in}}(t')\rangle & =\bar{n}_{\text{th}}\delta(t-t')\label{G1b_in2}
\end{align}
\end{subequations}

where $\bar{n}_{\text{th}}$ is the thermal occupancy of the phonon
bath at the mechanical resonance frequency.

\subsection{General solutions}

We write the four Langevin equations for the photon and phonon creation
and annihilation operators in the matrix form: $\dot{\mathbf{A}}=\mathbb{M}\mathbf{A}(t)+\mathbf{N}(t)$
where we have defined the vectors 
\[
\mathbf{A}=\begin{bmatrix}\hat{a}\\
\hat{a}^{\dag}\\
\hat{b}\\
\hat{b}^{\dag}
\end{bmatrix}\quad\text{and}\quad\mathbf{N}=\begin{bmatrix}\sqrt{\kappa}\hat{a}_{\text{in}}\\
\sqrt{\kappa}\hat{a}_{\text{in}}^{\dag}\\
\sqrt{\gamma}\hat{b}_{\text{in}}\\
\sqrt{\gamma}\hat{b}_{\text{in}}^{\dag}
\end{bmatrix}
\]
The matrix $\mathbb{M}$ is given by 
\[
\mathbb{M}=\begin{bmatrix}-\frac{\kappa}{2} & 0 & ig_{-} & ig_{+}\\
0 & -\frac{\kappa}{2} & -ig_{+} & -ig_{-}\\
ig_{-} & ig_{+} & -\frac{\gamma}{2} & 0\\
-ig_{+} & -ig_{-} & 0 & -\frac{\gamma}{2}
\end{bmatrix}
\]
To solve this system of first-order inhomogeneous linear differential
equations, we perform a change of basis to diagonalize the matrix
\[
\mathbb{M}\rightarrow\mathbb{M}^{D}=\mathbb{X}^{-1}\mathbb{MX}=\begin{bmatrix}\lambda_{1} & 0 & 0 & 0\\
0 & \lambda_{2} & 0 & 0\\
0 & 0 & \lambda_{3} & 0\\
0 & 0 & 0 & \lambda_{4}
\end{bmatrix}
\]
with the eigenvalues $\lambda_{1}=\lambda_{2}=-\frac{\kappa+\gamma}{4}-\frac{\sqrt{\Delta}}{2}$
and $\lambda_{3}=\lambda_{4}=-\frac{\kappa+\gamma}{4}+\frac{\sqrt{\Delta}}{2}$.
Here $\Delta=\frac{(\kappa-\gamma)^{2}}{4}-4(g_{-}^{2}-g_{+}^{2})$.

In the new basis $\mathbf{A}^{D}=\mathbb{X}^{-1}\mathbf{A}$ and $\mathbf{N}^{D}=\mathbb{X}^{-1}\mathbf{N}$
satisfy four uncoupled first-order differential equations 
\begin{equation}
\dot{\mathbf{A}}_{i}^{D}(t)=\lambda_{i}\mathbf{A}_{i}^{D}(t)+\mathbf{N}_{i}^{D}(t)\quad\text{for}\quad i=1...4
\end{equation}
where the noise operators play the role of driving terms. This can
easily be solved using the variation of the constant method to yield
\begin{equation}
\mathbf{A}_{i}^{D}(t)=\mathbf{A}_{i}^{D}(0)e^{\lambda_{i}t}+\int_{0}^{t}\mathbf{N}_{i}^{D}(t')e^{\lambda_{i}(t-t')}dt'\label{Diagonal_Solution_Components}
\end{equation}
We define the diagonal matrix 
\[
\mathbb{U}^{D}(t)=\delta_{i,j}\{e^{\lambda_{i}(t)}\}_{i,j}\quad i,j=1...4
\]
so that the solution writes 
\begin{equation}
\mathbf{A}^{D}(t)=\mathbb{U}^{D}(t)\mathbf{A}^{D}(0)+\int_{0}^{t}\mathbb{U}^{D}(t-t')\mathbf{N}^{D}(t')dt'\label{Diagonal_Solution_Vector}
\end{equation}
and transform back to the original basis 
\[
\mathbb{U}(t)\doteq\mathbb{XU}^{D}(t)\mathbb{X}^{-1}
\]
to obtain the time dependence of the original cavity operators 
\begin{equation}
\mathbf{A}(t)=\mathbb{U}(t)\mathbf{A}(0)+\int_{0}^{t}\mathbb{U}(t-t')\mathbf{N}(t')dt'\label{OriginalSolution}
\end{equation}

\subsection{Correlation functions}

We now proceed with the calculation of the higher-order moments ($G^{(1)},G^{(2)},G^{(3)}$)
of the optical and mechanical cavity operators. We define the covariance
matrix $\mathbb{G}^{(1)}(t_{1},t_{2})\doteq\langle\mathbf{A}(t_{1})\mathbf{A}^{T}(t_{2})\rangle$
with components $\mathbb{G}_{i,j}^{(1)}(t_{1},t_{2})=\langle\mathbf{A}_{i}(t_{1})\mathbf{A}_{j}(t_{2})\rangle$
and similarly for the noise operators $\mathbb{N}^{(1)}(t_{1},t_{2})\doteq\langle\mathbf{N}(t_{1})\mathbf{N}^{T}(t_{2})\rangle$
; $\mathbb{N}_{i,j}^{(1)}(t_{1},t_{2})=\langle\mathbf{N}_{i}(t_{1})\mathbf{N}_{j}(t_{2})\rangle=\delta(t_{1}-t_{2})\mathbb{N}_{i,j}^{(1)}$
where 
\[
\mathbb{N}_{i,j}^{(1)}=\begin{bmatrix}0 & \kappa & 0 & 0\\
0 & 0 & 0 & 0\\
0 & 0 & 0 & \gamma(\bar{n}_{\text{th}}+1)\\
0 & 0 & \gamma\bar{n}_{\text{th}} & 0
\end{bmatrix}
\]
Noting that the noise operators are stationary random fluctuations
with zero mean expectation values the following terms in $\mathbb{G}^{(1)}(t_{1},t_{2})$
vanish 
\begin{multline*}
\mathbb{U}(t_{1})\mathbf{A}(0)\times\int_{0}^{t_{2}}\mathbf{N}^{T}(t')\mathbb{U}^{T}(t_{2}-t')dt'=\\
\int_{0}^{t_{1}}\mathbb{U}(t_{1}-t')\mathbf{N}(t')dt'\times\mathbf{A}^{T}(0)\mathbb{U}^{T}(t_{2})=0
\end{multline*}
Therefore we obtain the expression for the first-order correlations
\begin{multline}
\mathbb{G}^{(1)}(t_{1},t_{2})=\langle\mathbb{U}(t_{1})\mathbf{A}(0)\mathbf{A}^{T}(0)\mathbb{U}^{T}(t_{2})+\\
\int_{0}^{t_{1}}dt_{1}'\int_{0}^{t_{2}}dt_{2}'\mathbb{U}(t_{1}-t_{1}')\mathbf{N}(t_{1}')\mathbf{N}^{T}(t_{2}')\mathbb{U}^{T}(t_{2}-t_{2}')\rangle\\
=\mathbb{U}(t_{1})\mathbb{G}^{(1)}(0,0)\mathbb{U}^{T}(t_{2})+\\
\int_{0}^{\text{min}(t_{1},t_{2})}\mathbb{U}(t_{1}-t')\mathbb{N}^{(1)}\mathbb{U}^{T}(t_{2}-t')dt'\label{G1_Matrix}
\end{multline}
Any component of $\mathbb{G}^{(2)}$ and $\mathbb{G}^{(3)}$ can then
be computed from the matrix $\mathbb{G}^{(1)}$ with the help of the
Gaussian moment-factoring theorem \cite{Ravazi_2009,Shapiro_94},
which states that the expectation value over a Gaussian state of any
four operators can be decomposed as a sum of ordered products 
\begin{multline}
\langle\widehat{a}_{1}\widehat{a}_{2}\widehat{a}_{3}\widehat{a}_{4}\rangle=\langle\widehat{a}_{1}\widehat{a}_{2}\rangle\langle\widehat{a}_{3}\widehat{a}_{4}\rangle+\langle\widehat{a}_{1}\widehat{a}_{3}\rangle\langle\widehat{a}_{2}\widehat{a}_{4}\rangle+\\
\langle\widehat{a}_{1}\widehat{a}_{4}\rangle\langle\widehat{a}_{2}\widehat{a}_{3}\rangle\label{Factorize}
\end{multline}

\subsection{Pulsed scheme and conditional $G^{(2)}$}

We consider the following experimental sequence: 
\begin{enumerate}
\item A red \textit{cooling} laser tuned at $\omega_{-}$ is used to optically
cool the mechanical resonator to an initial occupation number $\bar{n}_{0}$
below the bath average occupation $\bar{n}{}_{\text{th}}$. 
\item Immediately afterward a blue \textit{write} pulse at angular frequency
$\omega_{+}$ of duration $T_{\text{w}}$ (with $\kappa\ll T_{\text{w}}\ll\gamma\bar{n}_{\text{th}}$)
is applied to create a correlated photon/phonon pair via the parametric
down conversion interaction ($-\hbar g_{+}\hat{a}^{\dag}\hat{b}^{\dag}+h.c.$).
The photon of the pair is emitted at the central cavity frequency
$\omega_{c}$ and detected after spectral filtering to herald the
creation of a mechanical excitation. 
\item After a waiting time $T_{\text{off}}$ without laser excitation a
red \textit{readout} pulse at angular frequency $\omega_{-}$ is used
to map the mechanical state onto the optical cavity mode at $\omega_{c}$
via the beam-splitter interaction ($-\hbar g_{+}\hat{a}\hat{b}^{\dag}+h.c.$).
The photons leaking out of the cavity are sent to a Hanbury-Brown
and Twiss setup to measure their second-order correlation function
$G^{(2)}$, conditional on the detection of a heralding photon in
the previous step (in effect a third-order correlation measurements
$G^{(3)}$). 
\end{enumerate}
We compute the normalized second-order correlation between photons
detected during the readout pulse at times $t_{\text{r}}$ and $t_{\text{r}}+\tau$
(with respect to the beginning of the red pulse), conditional on the
detection of a photon emitted at time $t_{\text{w}}$ (with respect
to the beginning of the blue pulse) during the excitation pulse 
\begin{equation}
g_{\text{cond}}^{(2)}(t_{\text{r}},t_{\text{r}}+\tau\vert t_{\text{w}})=\frac{G_{\text{cond}}^{(2)}(t_{\text{r}},t_{\text{r}}+\tau\vert t_{\text{w}})}{G_{\text{cond}}^{(1)}(t_{\text{r}}\vert t_{\text{w}})G_{\text{cond}}^{(1)}(t_{\text{r}}+\tau\vert t_{\text{w}})}\label{Cond_g2}
\end{equation}
Following Ref. \citep{Ravazi_2009} we express the conditional correlations
(i.e. the post-measurement expectation values) \begin{subequations}
\begin{align}
G_{\text{cond}}^{(1)}(t_{\text{r}}\vert t_{\text{w}}) & =\frac{G^{(2)}(t_{\text{r}},t_{\text{w}})}{G^{(1)}(t_{\text{w}},t_{\text{w}})}\\
G_{\text{cond}}^{(2)}(t_{\text{r}},t_{\text{r}}+\tau\vert t_{\text{w}}) & =\frac{G^{(3)}(t_{\text{r}},t_{\text{r}}+\tau,t_{\text{w}})}{G^{(1)}(t_{\text{w}},t_{\text{w}})}
\end{align}
\end{subequations} with the functions $G^{(n)}$ standing for correlations
between any operators, in particular for the photon cavity mode 
\begin{align*}
G_{a}^{(1)}(t) & =\langle\hat{a}^{\dag}(t)\hat{a}(t)\rangle\\
G_{a}^{(2)}(t,t') & =\langle\hat{a}^{\dag}(t')\hat{a}^{\dag}(t)\hat{a}(t)\hat{a}(t')\rangle\\
G_{a}^{(3)}(t,t',t'') & =\langle\hat{a}^{\dag}(t'')\hat{a}^{\dag}(t')\langle\hat{a}^{\dag}(t)\hat{a}(t)\hat{a}(t')\hat{a}(t'')\rangle
\end{align*}
All these quantities are computed from equations (\ref{G1_Matrix})
and (\ref{Factorize}) using \textit{Mathematica}. \\

\section{Explicit conditional state}

In order to derive analytic expressions for the conditional state
of the mechanical mode and the photon correlation function we use
the fact that in typical experimental scenarios $g_{\pm}\ll\kappa$
(weak coupling) which allows for adiabatic elimination of the cavity
mode. We also neglect the decay of the mechanical excitations, which
is a valid approximation as long as the complete pulse sequence occurs
within a duration shorter than the thermal decoherence rate $\sim\gamma n_{\text{th}}$.

\subsection{Blue write pulse}

Under these conditions the Langevin equations (\ref{Lang_a},\ref{Lang_b})
during the blue \textit{write} pulse (operators are labeled with the
subscript $\text{w}$ during this step) become \begin{subequations}
\begin{align}
\frac{d\hat{a}_{\text{w}}}{dt} & =ig_{+}\hat{b}_{\text{w}}^{\dag}-\frac{\kappa}{2}\hat{a}_{\text{w}}+\sqrt{\kappa}\hat{a}_{\text{w,in}}\\
\frac{d\hat{b}_{\text{w}}}{dt} & =ig_{+}\hat{a}_{\text{w}}^{\dag}
\end{align}
\end{subequations} After adiabatic elimination: $\hat{a}_{\text{w}}(t)\simeq\frac{2}{\kappa}(ig_{+}\hat{b}_{\text{w}}^{\dag}+\sqrt{\kappa}\hat{a}_{\text{w,in}})$
and using the input/output relations \cite{Gardiner_1985}: $\hat{a}_{\text{w,out}}=-\hat{a}_{\text{w,in}}+\sqrt{\kappa}\hat{a}_{\text{w}}$
, we obtain \begin{subequations} 
\begin{align}
\hat{a}_{\text{w,out}} & =\hat{a}_{\text{w,in}}+i\sqrt{2\tilde{g}_{\text{w}}}\hat{b}_{\text{w}}^{\dag}\label{Lang_wa}\\
\frac{d\hat{b}_{\text{w}}}{dt} & =\tilde{g}_{\text{w}}\hat{b}_{\text{w}}+i\sqrt{2\tilde{g}_{\text{w}}}\hat{a}_{\text{w,in}}^{\dag}\label{Lang_wb}
\end{align}
\end{subequations} where we have defined $\tilde{g}_{\text{w}}\equiv\frac{2g_{+}^{2}}{\kappa}$.
We follow Hofer \textit{et al.} \cite{Hofer_2011} and introduce the
temporal optical modes 
\begin{align*}
\hat{A}_{\text{w,in}}(t) & =\sqrt{\frac{2\tilde{g}_{\text{w}}}{1-e^{-2\tilde{g}_{\text{w}}t}}}\int_{0}^{t}dt'e^{-\tilde{g}_{\text{w}}t'}\hat{a}_{\text{w,in}}(t')\\
\hat{A}_{\text{w,out}}(t) & =\sqrt{\frac{2\tilde{g}_{\text{w}}}{e^{2\tilde{g}_{\text{w}}t}-1}}\int_{0}^{t}dt'e^{\tilde{g}_{\text{w}}t'}\hat{a}_{\text{w,out}}(t')
\end{align*}
Integrating equations (\ref{Lang_wa}-\ref{Lang_wb}) then leads to
the simple results \begin{subequations} 
\begin{align}
\hat{A}_{\text{w,out}}(t) & =e^{\tilde{g}_{\text{w}}t}\hat{A}_{\text{w,in}}(t)+i\sqrt{e^{2\tilde{g}_{\text{w}}t}-1}\hat{b}_{0}^{\dag}\label{Sol_wa}\\
\hat{b}_{\text{w}}(t) & =e^{\tilde{g}_{\text{w}}t}\hat{b}_{0}+i\sqrt{e^{2\tilde{g}_{\text{w}}t}-1}\hat{A}_{\text{w,in}}^{\dag}(t)\label{Sol_wb}
\end{align}
\end{subequations} The analogy with the two-mode squeezing interaction
occuring in optical parametric down-conversion becomes obvious if
we introduce the squeezing parameter $r$ and identify formally $e^{\tilde{g}_{\text{w}}t}\equiv\cosh{r}$;
$\sqrt{e^{2\tilde{g}_{\text{w}}t}-1}=\sqrt{\cosh^{2}{r}-1}=\sinh{r}$.
Through the solutions (\ref{Sol_wa},\ref{Sol_wb}), it is possible
to extract the propagator $U(t)$ which satisfies $\hat{A}_{\text{w,out}}(t)=U^{\dagger}(t)\hat{A}_{\text{w,in}}(t)U(t)$
and $\hat{b}_{\text{w}}(t)=U(t)^{\dagger}\hat{b}_{\text{0}}U(t)$.
Its explicit expression is

\begin{multline}
\hat{U}(t)=e^{i\tanh r\hat{A}_{\text{w,in}}^{\dag}\hat{b}_{\text{0}}^{\dag}}\\
\cosh{r}^{(-1-\hat{A}_{\text{w,in}}^{\dag}\hat{A}_{\text{w,in}}-\hat{b}_{\text{0}}^{\dag}\hat{b}_{\text{0}})}e^{-i\tanh r\hat{A}_{\text{w,in}}\hat{b}_{\text{0}}}
\end{multline}
 An initial state $\vert0_{A},0_{b}\rangle$ thus evolves towards
\begin{equation}
\vert\Psi(T_{\text{w}})\rangle=e^{-\tilde{g}_{\text{w}}T_{\text{w}}}\sum_{n=0}^{\infty}\frac{i^{n}}{n!}(1-e^{-2\tilde{g}_{\text{w}}T_{\text{w}}})^{n/2}\vert n_{A},n_{b}\rangle
\end{equation}
at the end of a write pulse of duration $T_{\text{w}}$. We find that
the conditional state of the phonon mode upon detection of a single
photon is indeed a Fock state $\vert1_{b}\rangle$. In the realistic
case where non photon number resolving detectors are used, this remains
true to a good approximation as long as the probability for creating
a photon/phonon pair is sufficiently low, i.e. for $(1-e^{-2\tilde{g}_{\text{w}}T_{\text{w}}})\ll1$,
so that the probability for creating more than one photon-photon pair
is negligible.

\subsection{Red readout pulse}

We now consider the readout process, during which a laser tuned to
the red (lower) mechanical sideband is used to swap the mechanical
and optical states through the beam-splitter interaction. The simplified
Langevin equations for this step are \begin{subequations} 
\begin{align}
\frac{d\hat{a}_{\text{r}}}{dt} & =ig_{-}\hat{b}_{\text{r}}-\frac{\kappa}{2}\hat{a}_{\text{r}}+\sqrt{\kappa}\hat{a}_{\text{r,in}}\\
\frac{d\hat{b}_{\text{r}}}{dt} & =ig_{-}\hat{a}_{\text{r}}
\end{align}
\end{subequations} Following the same procedure as before and adiabatically
eliminating the optical cavity evolution we obtain \begin{subequations}
\begin{align}
\hat{a}_{\text{r,out}} & =\hat{a}_{\text{r,in}}+i\sqrt{2\tilde{g}_{\text{r}}}\hat{b}_{\text{r}}^{\dag}\label{Lang_r}\\
\frac{d\hat{b}_{\text{r}}}{dt} & =-\tilde{g}_{\text{r}}\hat{b}_{\text{r}}+i\sqrt{2\tilde{g}_{\text{r}}}\hat{a}_{\text{r,in}}^{\dag}\label{Lang_r2}
\end{align}
\end{subequations} where $\tilde{g}_{\text{r}}\equiv\frac{2g_{-}^{2}}{\kappa}$.
We define the readout temporal modes as \citep{Hofer_2011} 
\begin{align*}
\hat{A}_{\text{r,in}}(t) & =\sqrt{\frac{2\tilde{g}_{\text{r}}}{e^{2\tilde{g}_{\text{r}}t}-1}}\int_{0}^{t}dt'e^{\tilde{g}_{\text{r}}t'}\hat{a}_{\text{r,in}}(t')\\
\hat{A}_{\text{r,out}}(t) & =\sqrt{\frac{2G_{\text{r}}}{1-e^{-2G_{\text{r}}t}}}\int_{0}^{t}dt'e^{-\tilde{g}_{\text{r}}t'}\hat{a}_{\text{r,out}}(t')
\end{align*}
which leads to the simple expression for the solution of (\ref{Lang_r}-\ref{Lang_r2})
at a time $T_{\text{r}}$ after the beginning of the readout pulse
\begin{subequations} 
\begin{align}
\hat{A}_{\text{r,out}}(T_{\text{r}}) & =e^{-\tilde{g}_{\text{r}}T_{\text{r}}}\hat{A}_{\text{r,in}}(T_{\text{r}})+i\sqrt{1-e^{-2\tilde{g}_{\text{r}}T_{\text{r}}}}\hat{b}_{\text{r}}(0)\label{Arout}\\
\hat{b}_{\text{r}}(T_{\text{r}}) & =e^{-\tilde{g}_{\text{r}}T_{\text{r}}}\hat{b}_{\text{r}}(0)+i\sqrt{1-e^{-2\tilde{g}_{\text{r}}T_{\text{r}}}}\hat{A}_{\text{r,in}}(T_{\text{r}})\label{Sol_rb}
\end{align}
\end{subequations}

\subsection{Conditional phonon state and photon correlations}

We assume the mechanical mode to be initially in a thermal state with
average phonon occupancy $\bar{n}_{0}$, characterized by the density
matrix 
\[
\rho_{b}(0)=(1-p)\sum_{n\geq0}p^{n}\vert n\rangle\langle n\vert\quad;\quad p\equiv\frac{\bar{n}_{0}}{1+\bar{n}_{0}}
\]
The phonon average occupancy is indeed recovered by the usual trace
formula with the operator $\hat{b}^{\dag}\hat{b}$ 
\begin{multline*}
tr_{b}(\hat{b}^{\dag}\hat{b}\rho_{b})=(1-p)\sum_{n\geq0}p^{n}n=(1-p)\frac{p}{(1-p)^{2}}\\
=\frac{p}{1-p}=\bar{n}_{0}
\end{multline*}
The density matrix of the coupled optomechanical system just before
the \textit{write} pulse is taken to be in a product state 
\[
\rho_{A,b}(0)=\vert0_{A}\rangle\langle0_{A}\vert\otimes\rho_{b}(0)=(1-p)\sum_{n\geq0}p^{n}\vert0,n\rangle\langle0,n\vert
\]

During the blue \textit{write} pulse, the thermal excitations of the
mechanics act as a seed for the parametric down-conversion process.
The average number of photons emitted into the cavity mode during
the blue pulse is $\bar{n}_{\text{\text{w}}}=tr(\hat{A}_{\text{w,out}}^{\dag}\hat{A}_{\text{w,out}}\rho_{A,b})$
(the trace is over both modes), which we estimate using the solution
found in (\ref{Sol_wa}) for a pulse of duration $T_{\text{w}}$ 
\begin{align*}
N_{\text{herald}} & =tr(\vert0_{A}\rangle\langle0_{A}\vert\otimes(e^{2\tilde{g}_{\text{w}}T_{\text{w}}}-1)\hat{b}\hat{b}^{\dag}\rho_{b})\\
 & =(e^{2\tilde{g}_{\text{w}}T_{\text{w}}}-1)tr_{b}((1-p)\sum_{n\geq0}p^{n}(n+1)\vert n+1\rangle\langle n+1\vert)\\
 & =(e^{2\tilde{g}_{\text{w}}T_{\text{w}}}-1)(1-p)\sum_{n\geq0}p^{n}(n+1)\\
 & =(e^{2\tilde{g}_{\text{w}}T_{\text{w}}}-1)\frac{1}{1-p}\\
 & =(e^{2\tilde{g}_{\text{w}}T_{\text{w}}}-1)(\bar{n}_{0}+1)
\end{align*}
As expected, the factor $(\bar{n}_{0}+1)$ corresponds to the stimulated
emission of photons by the presence of thermal phonons.

The \textit{unnormalized} conditional state of the mechanics upon
detection of a single Stokes photon is given by $\rho_{b}^{\text{*cond}}=tr_{A}(\vert1_{A}\rangle\langle1_{A}\vert\otimes\mathbb{I}_{b})U\rho_{A,b}(0)U^{\dagger}$,
where 
\begin{multline*}
U\rho_{A,b}(0)U^{\dagger}=(1-p)\sum_{n\geq0}p^{n}\cosh{r}^{-2-2n}e^{i\tanh{r}\hat{A}_{\text{in}}^{\dag}\hat{b}_{0}^{\dag}}\\
\vert0,n\rangle\langle0,n\vert e^{-i\tanh{r}\hat{A}_{\text{in}}\hat{b}_{0}}
\end{multline*}
so that 
\begin{multline*}
\rho_{b}^{\text{*cond}}=(1-p)\sum_{n\geq0}p^{n}\cosh{r}^{-2-2n}\tanh^{2}{r}(n+1)\\
\vert n+1\rangle\langle n+1\vert
\end{multline*}
The probability of detecting a single photon is given by the norm
of this state 
\begin{align*}
tr(\rho_{b}^{\text{*cond}}) & =(1-p)\tanh^{2}{r}\cosh^{-2}{r}\sum_{n\geq0}(p\cosh^{-2}{r})^{n}(n+1)\\
 & =(1-p)\frac{\tanh^{2}{r}}{\cosh^{2}{r}(1-p\cosh^{-2}{r})^{2}}
\end{align*}
Recalling that $\cosh{r}=e^{\tilde{g}_{\text{w}}T_{\text{w}}}$ and
defining $\overline{p}=pe^{-\tilde{g}_{\text{w}}T_{\text{w}}}$ we
obtain the normalized conditional state 
\[
\rho_{b}^{\text{cond}}=(1-\overline{p})^{2}\sum_{n\geq0}\overline{p}^{n}(n+1)\vert n+1\rangle\langle n+1\vert
\]
For small enough gain ($\tilde{g}_{\text{w}}T_{\text{w}}\ll1$) and
an initial resonator close to its ground state ($p\ll1$) the dominant
term in the conditional state is indeed the single phonon Fock state
$\vert1\rangle$. The bi-phonon component is smaller by a factor $1/\overline{p}$.

We can now compute the conditional (heralded) second order correlation
function of the Anti-Stokes photons during the \textit{readout} pulse,
$g_{\text{cond}}^{(2)}(0)=\frac{\langle\hat{A}_{\text{r,out}}^{\dag2}\hat{A}_{\text{r,out}}^{2}\rangle_{c}}{\langle\hat{A}_{\text{r,out}}^{\dag}\hat{A}_{\text{r,out}}\rangle_{c}^{2}}$
where the expectation value is taken on the post-selected state at
the beginning of the readout pulse, $\vert0_{A_{\text{r,in}}}\rangle\langle0_{A_{\text{r,in}}}\vert\otimes\rho_{b}^{\text{cond}}$.
From (\ref{Arout}) we obtain for the numerator and denominator:

\begin{align*}
\langle\hat{A}_{\text{r,out}}^{\dag2}\hat{A}_{\text{r,out}}^{2}\rangle_{c} & =(1-e^{-2\tilde{g}_{\text{r}}T_{\text{r}}})^{2}(1-\overline{p})^{2}\sum_{n\geq0}\overline{p}^{n}(n+1)^{2}n\\
 & =(1-e^{-2\tilde{g}_{\text{r}}T_{\text{r}}})^{2}\frac{2\overline{p}(2+\overline{p})}{(1-\overline{p})^{2}}\\
\\
\langle\hat{A}_{\text{r,out}}^{\dag}\hat{A}_{\text{r,out}}\rangle_{c} & =(1-e^{-2\tilde{g}_{\text{r}}T_{\text{r}}})(1-\overline{p})^{2}\sum_{n\geq0}\overline{p}^{n}(n+1)^{2}\\
 & =(1-e^{-2\tilde{g}_{\text{r}}T_{\text{r}}})\frac{1+\overline{p}}{1-\overline{p}}
\end{align*}
And therefore 
\begin{align}
g_{\text{cond}}^{(2)}(0) & =\frac{2\overline{p}(2+\overline{p})}{(1+\overline{p})^{2}}\\
 & \approx4\bar{n}_{0}\quad\text{for}\quad\bar{n}_{0},\tilde{g}_{\text{w}}T_{\text{w}}\ll1
\end{align}

\section{Ground state cooling}

\begin{figure}
\includegraphics[scale=1.2]{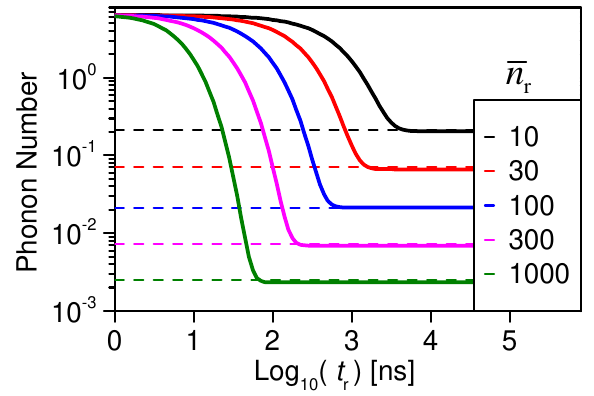} \protect\caption{\label{figS1} Mean phonon number in the resonator mode $\langle\hat{b}^{\dag}\hat{b}\rangle$
as a function of time starting from an initial occupancy $\bar{n}_{0}=\bar{n}_{\text{th}}=6.4$
(corresponding the parameters in the main text, $T_{\text{bath}}=1.6$~K
and $\Omega_{m}=5.1$~GHz) for increasing cooling laser power, expressed
as intracavity photon number $N_{\text{r}}$ at the lower sideband
frequency. Solid lines: calculations from the Langevin equations neglecting
counter-rotating terms, i.e. without quantum backaction (see section~\ref{Sec1}).
Dashed lines: expected final occupancy in the resolved sideband regime
when accounting for quantum backaction. For this range of cooling
powers quantum backaction effects are negligible. }
\end{figure}

In the main text we consider that the mechanical oscillator is optically
cooled to an initial occupancy $\bar{n}_{0}=10^{-2}$ whereas the
bath temperature $T_{\text{bath}}=1.6$~K corresponds to an average
phonon number $\bar{n}_{\text{th}}=6.4$ at 5.1~GHz. Here we justify
the feasibility of this scenario and show that re-cooling can be achieved
within $\sim100$~ns after each \textit{write/readout} sequence,
setting an upper bound of 10~MHz to the repetition rate of the experiment.
In Fig.~\ref{figS1} we show the time evolution of the phonon population
$\langle\hat{b}^{\dag}\hat{b}\rangle$ calculated from the linearized
Langevin equations (scetion \ref{Sec1}). When the oscillator is initially
in equilibrium with the bath, we find that a final occupancy below
$10^{-2}$ can be reached for $\bar{n}_{\text{r}}\gtrsim300$, which
is still well in the weak-coupling regime. Since in our calculations
we neglect the counter-rotating terms at $\pm2\Omega_{m}$ there is
no limiting quantum backaction. To check that this effect is negligible,
we also plot in Fig.~\ref{figS1} the formula derived elsewhere \cite{
Liu_2013b}
(under the resolved-sideband approximation) 
\begin{equation}
\langle\hat{b}^{\dag}\hat{b}\rangle_{\text{fin}}=\frac{\gamma(4g_{0}^{2}\bar{n}_{\text{r}}+\kappa^{2})}{4g_{0}^{2}\bar{n}_{\text{r}}(\kappa+\gamma)}\bar{n}_{\text{th}}+\frac{\kappa^{2}+8g_{0}^{2}\bar{n}_{\text{r}}}{16(\Omega_{m}^{2}-4g_{0}^{2}\bar{n}_{\text{r}})}\label{Nfinal}
\end{equation}
where the second term accounts for the quantum backaction limit to
cooling. It can be seen that the deviation due to quantum backaction
is indeed very small even for $\bar{n}_{\text{r}}=1000$. Since $\bar{n}_{\text{r}}=\vert\frac{-is_{-}}{i\Omega_{m}+\kappa/2}\vert^{2}=\frac{\kappa P_{-}}{\hbar\omega_{c}(\Omega_{m}^{2}+\kappa^{2}/4)}$
this value of $\bar{n}_{\text{r}}$ corresponds to 150~$\mu$W of
external laser power, which emphasizes the need for good thermalization
of the device in the cryostat to avoid heating the bath due to spurious
light absorption.

\end{document}